\documentclass[aps,prl,reprint,nobibnotes,a4paper,amsmath,
nobalancelastpage,showkeys]{revtex4-2}

\begin{document}
\title{General expression for three-loop \mbox{$\beta$-functions} of $\cal N =$ 1 supersymmetric theories with multiple gauge couplings regularized by higher covariant derivatives}
\author{O.\,V.\,Haneychuk}
\affiliation{Moscow State University, 119991, Leninskie Gory, 1, Moscow, Russia}
\email{oleshan91@gmail.com}

\begin{abstract}
For $\cal N =$ 1 supersymmetric theories with multiple gauge couplings regularized by higher covariant derivatives, a general expression for three-loop gauge $\beta$-functions is obtained. For this purpose, using general statements about the validity of the NSVZ relations, a result is constructed for the \mbox{$\beta$-functions} defined in terms of the bare couplings. It is demonstrated that in the particular case of MSSM, it precisely reproduces the known expressions found earlier in another way. For the case where Yukawa couplings are absent, similar expressions for three-loop $\beta$-functions are also obtained in terms of the renormalized couplings in an arbitrary subtraction scheme and, in particular, in the $\overline{\mbox{DR}}$ scheme.
\end{abstract}

\maketitle
In elementary particle physics, the hypothesis that the Standard Model is a low-energy limit of some Grand Unified Theory with a larger gauge group that includes the gauge group of the Standard Model as a subgroup has been explored for a long time. According to this assumption, all three gauge couplings of the Standard Model should take the same value at the unification scale. In the Standard Model such a scale is absent, but in its \mbox{$\cal N$ = 1} supersymmetric extensions, due to the presence of superpartners, the unification of couplings actually takes place \cite{Ellis1991,Amaldi1991,Langacker1991}. An interesting feature of $\cal N$ = 1 supersymmetric gauge theories is the existence of a relation between the evolution of the gauge couplings, determined by the corresponding $\beta$-functions, and the renormalization of the matter superfields. In the case where the gauge group is a direct product of $n$ subgroups $G_K$, each of which is either simple or $U(1)$,

\begin{equation}
G = \prod\limits_K G_K = G_1\times G_2\times \ldots \times G_n,
\end{equation}
\noindent
there are $n$ gauge couplings $\alpha_K$ and, therefore, $n$ gauge $\beta$-functions. In our notation, capital Latin letters numerate the subgroups of the gauge group. 

The renormalization constants $Z$ of the matter superfields are defined by the equation $\phi = \sqrt Z\phi_{R}$, where $\phi_R$ denotes the renormalized superfield. The anomalous dimensions are constructed based on the renormalization constants. They can be defined both in terms of the bare gauge and Yukawa couplings $\alpha_0$ and $\lambda_0$, and in terms of the renormalized couplings $\alpha$ and $\lambda$. In terms of the bare couplings, gauge \mbox{$\beta$-functions} and anomalous dimensions are respectively defined as follows:

\begin{equation}
\beta_K(\alpha_0,\lambda_0) \equiv \frac{d\alpha_{0K}}{d\ln\Lambda}\bigg|_{\alpha,\lambda=\text{const}},\label{beta_0}
\end{equation}
\begin{equation}
{\gamma}(\alpha_0,\lambda_0) \equiv -\frac{d\ln {Z}}{d\ln\Lambda}\bigg|_{\alpha,\lambda=\text{const}},\label{gamma_0}
\end{equation}
\noindent
where $\Lambda$ is the regularization parameter, which is in fact an ultraviolet cut-off. The renormalization group functions (RGF) (\ref{beta_0}) and (\ref{gamma_0}) depend on regularization parameters but do not depend on a subtraction scheme. However, the $\beta$-functions and anomalous dimensions defined in terms of the renormalized couplings are more widely used:

\begin{equation}
\widetilde\beta_K(\alpha,\lambda) \equiv \frac{d\alpha_{K}}{d\ln\mu}\bigg|_{\alpha_0,\lambda_0=\text{const}},\label{beta_r}
\end{equation}
\begin{equation}
\widetilde{\gamma}(\alpha,\lambda) \equiv \frac{d\ln {Z}}{d\ln\mu}\bigg|_{\alpha_0,\lambda_0=\text{const}},\label{gamma_r}
\end{equation}
\noindent
where $\mu$ is the renormalization point. The \mbox{$\beta$-functions} (\ref{beta_r}) and the anomalous dimensions (\ref{gamma_r}) depend on both a regularization and a renormalization prescription. This property is important to keep in mind because it affects the validity of important relations inherent in \mbox{$\cal N$ = 1} supersymmetric theories and known as the NSVZ relations \cite{Novikov1983,Jones1983,Novikov1986,Shifman1986}.
These relations express the $\beta$-functions in terms of the anomalous dimensions of the matter superfields found in all previous loops. Therefore, it is possible to calculate $\beta$-functions with the help of the calculation of the anomalous dimensions of matter superfields in lower orders of the perturbation theory. For this purpose, it is important to use a renormalization prescription in which the NSVZ relations are satisfied. For example, they are satisfied in all orders if a theory is regularized by higher covariant derivatives \cite{Slavnov1971,Slavnov1972} and RGFs are defined in terms of the bare couplings \cite{kataev2013, Stepanyantz2020}. There are renormalization prescriptions under which the NSVZ relations remain valid for RGFs defined in terms of the renormalized couplings as well. They include, for example, the HD+MSL scheme, in which a theory is regularized by higher covariant derivatives, and the renormalization of couplings and matter superfields includes only powers of $\ln\Lambda/\mu$. The most widely used in supersymmetric theories, however, is the $\overline{\mbox{DR}}$ scheme \cite{siegel1979}, for which the NSVZ relations are not valid in those orders of the perturbation theory where the scheme dependence manifests itself, but can be restored with the help of a specially tuned finite renormalization \cite{Jack1996_1,Jack1996_2}. Here we will use the higher covariant derivative regularization which allows to avoid the above mentioned difficulties. It should be noted that the ordinary addition of a term with higher covariant derivatives does not eliminate divergences in the one-loop approximation. That is why the considered regularization also includes the insertion of Pauli--Villars determinants \cite{Slavnov1977}, which cancel divergences arising from the one-loop diagrams. One-loop divergences from the gauge superfields and ghosts are eliminated by introducing the Pauli--Villars superfields $\varphi_{1,K}$, $\varphi_{2,K}$, $\varphi_{3,K}$ with the masses $M_{\varphi,K}=a_{\varphi,K}\Lambda$, and divergences arising from the one-loop diagrams of matter superfields are removed by introducing the Pauli--Villars superfields with the masses $M_{K}=a_{K}\Lambda$.

The NSVZ relations are valid for theories with multiple gauge couplings as well \cite{Shifman1996}. In this case, the NSVZ relations can be written as \cite{Korneev2021}

\begin{eqnarray}
\frac{\beta_K(\alpha_0\mbox{,}\lambda_0)}{{\alpha_{0K}}^2}&=&-\frac{1}{2\pi(1-C_2(G_K)\alpha_{0K}/2\pi)}\times \label{NSVZ} \\
&\times &\Big[3C_2(G_K)
-\sum_aT_{aK}(1-{\gamma_a}{}^a(\alpha_0\mbox{,}\lambda_0)) \Big],\nonumber
\end{eqnarray}
\noindent
where the index $a$ numerates the  matter superfields $\phi_a$ that lie in the irreducible representations $R_{aK}$ of the simple subgroups $G_K$ and have charges $q_{aK}$ with respect to the subgroups $U(1)$. The complete set of indices for the matter superfields can be represented as $(a;i_1,i_2,\ldots,i_n)\equiv (a;i_a)$. We denote the anomalous dimensions of the superfields $\phi_a$ by $\gamma_a{}^b$, and determine the group factors by the equalities

\begin{eqnarray}
&& C_2(G_K)\delta^{A_K B_K} = f^{A_K C_KD_K}f^{B_K C_KD_K}; \qquad\nonumber\\
&& T_K(R_{aK})\delta^{A_K B_K} = {(T_a^{A_K}T_a^{B_K})_{i_K}}^{i_K};
\end{eqnarray}

\begin{equation}
\begin{split}
& T_{aK} =\\
& \begin{cases}
   {\delta_{i_1}}^{i_1} \ldots  {\delta_{i_{K-1}}}^{i_{K-1}}T_K(R_{ak}){\delta_{i_{K+1}}}^{i_{K+1}} \ldots {\delta_{i_n}}^{i_n}, \\
   \text{\hspace{32mm}  $G_K$ is simple;}\\
   {\delta_{i_1}}^{i_1} \ldots {\delta_{i_{K-1}}}^{i_{K-1}}q^2_{aK}{\delta_{i_{K+1}}}^{i_{K+1}} \ldots {\delta_{i_n}}^{i_n}, \\
  \text{\hspace{32mm}  $G_K = U(1)$},
 \end{cases}
 \end{split} \nonumber
\end{equation}
\noindent
where $T_a^{A_K}$ for the simple subgroups $G_K$ are the generators in the representation in which the superfields $\phi_a$ lie, and for subgroups $U(1)$ are the corresponding charges $q_a$.

In the case of using the higher covariant derivative regularization, the relations (\ref{NSVZ})  can be used to find $L$-loop gauge $\beta$-functions for $\cal N =$ 1 supersymmetric theories with multiple couplings, expressed in terms of the bare couplings. To do this, it is necessary to know the general expression for the anomalous dimensions of the matter superfields ${\gamma_a}^b(\alpha_0\mbox{,}\lambda_0)$ in all previous orders. For $\cal N =$ 1 supersymmetric Yang--Mills theory with a simple gauge group, a general expression for $\gamma_a{}^b$ in the two-loop approximation was found in \cite{Kazantsev2020}. In \cite{Stepanyantz2022} this expression was generalized to the case of theories with multiple couplings regularized by higher covariant derivatives,

\begin{eqnarray}
&&{{\gamma_a}^b}(\alpha_0 \mbox{,}\lambda_0) \equiv
 -\frac{d \ln {Z_a}^b}{d \ln \Lambda}\Big|_{\alpha,\lambda=\mbox{\scriptsize const}} \nonumber \\
&& = -\sum\limits_{K} \frac{\alpha_{0K}}{\pi}C(R_{aK}){\delta_a}^b
+\frac{1}{4\pi^2}{(\lambda_{0}^*\lambda_0)_a}^b \nonumber \\
&& +\sum\limits_{KL}\frac{\alpha_{0K}\alpha_{0L}}{2\pi^2}C(R_{aK})C(R_{aL}){\delta_a}^b  \nonumber\\
&& - \sum\limits_{K}\frac{\alpha_{0K}^2}{2\pi^2}C(R_{aK})\Big[3C_2(G_K)\ln a_{\varphi\mbox{,}K} - \sum\limits_{c}T_{cK}\ln a_{K} \nonumber\\
&& - Q_K\Big(1 +\frac{A}{2}\Big)\Big]{\delta_a}^b   - \sum\limits_{K}\frac{\alpha_{0K}}{8\pi^3}{(\lambda_{0}^*\lambda_0)_a}^bC(R_{aK})\times\nonumber \\
&&\times\Big(1-B+A\Big) + \sum\limits_{K}\frac{\alpha_{0K}}{4\pi^3}{(\lambda_{0}^*C_K\lambda_0)_a}^b\Big(1+B-A\Big) \nonumber\\
&& - \frac{1}{16\pi^4}{(\lambda_{0}^*[\lambda_{0}^*\lambda_0]\lambda_0)_a}^b  +O\Big(\alpha_0^3 \mbox{,}\alpha_0^2\lambda_0^2 \mbox{,}\alpha_0\lambda_0^4\mbox{,}\lambda_0^6\Big),
\label{gamma}
\end{eqnarray}
\noindent
where $Q_K = \sum\limits_{a}T_{aK} - 3C_2(G_K)$. This expression contains the regularization parameters $a_{\varphi,K}$ and $a_K$, which are the ratios of the masses of the Pauli--Villars superfields to the parameter $\Lambda$, as well as parameters

\begin{eqnarray}
&& A = \int\limits_0^\infty dx\,\ln x \frac{d}{dx}\frac{1}{R(x)};\qquad\nonumber\\
&& B = \int\limits_0^\infty dx\,\ln x \frac{d}{dx}\frac{1}{F^2(x)},
\end{eqnarray}
\noindent
where the functions $R(x)$ and $F(x)$ determine the regularizing terms with higher derivatives in the gauge part of the action and in the kinetic term for the chiral matter superfields, respectively. A detailed description of these issues can be found in \cite{Korneev2021}. Also in the equation (\ref{gamma}) the notations 

\begin{eqnarray}
&& C(R_{aK}) \equiv T_a^{A_K}T_a^{A_K};\nonumber\\
&& {(T_a^{A_K}T_a^{A_K})_{i_K}}^{j_K}=C(R_{aK}){\delta_{i_K}}^{j_K}; \nonumber\\
&& {(\lambda_{0}^*\lambda_0)_a}^b{\delta_{i_a}}^{j_b}=\sum\limits_{cd}\lambda_{0i_am_cn_d}^*\lambda_0^{j_bm_cn_d}; \nonumber\\
&& {(\lambda_{0}^*C_K\lambda_0)_a}^b{\delta_{i_a}}^{j_b}=\sum\limits_{cd}\lambda_{0i_am_cn_d}^*C(R_{dK})\lambda_0^{j_bm_cn_d}; \nonumber\\
&& {(\lambda_{0}^*[\lambda_{0}^*\lambda_0]\lambda_0)_a}^b{\delta_{i_a}}^{j_b}= \sum\limits_{cdefg}\lambda_{0i_ak_el_f}^*\lambda_0^{j_bk_ep_g}\nonumber\\
&& \times \lambda_{0p_gm_cn_d}^*\lambda_0^{l_fm_cn_d}
\end{eqnarray}
\noindent
are used. Under the used regularization, as a result of substituting the anomalous dimensions (\ref{gamma}) into the relations (\ref{NSVZ}), one can obtain the general expression for three-loop $\beta$-functions in \mbox{$\cal N =$ 1} supersymmetric theories with several couplings,

\begin{eqnarray}
&& \frac{\beta_K(\alpha_0,\lambda_0)}{\alpha_{0K}^2}=-\frac{1}{2\pi}\Big\{-Q_K - \frac{\alpha_{0K}}{2\pi}C_2(G_K)Q_K \nonumber\\
&& -\sum\limits_{a}\sum\limits_{L}\frac{\alpha_{0L}}{\pi}T_{aK}C(R_{aL})  +\frac{1}{4\pi^2}\sum\limits_{abc}T_{aK}\lambda_{0i_a m_b n_c}^* \nonumber\\
&& \times\lambda_0^{i_a m_b n_c} - \sum\limits_{a}\sum\limits_{L}\frac{\alpha_{0K}\alpha_{0L}}{2\pi^2}T_{aK}C_2(G_K)C(R_{aL})\nonumber\\
&& -\frac{\alpha_{0K}^2}{4\pi^2}C_2^2(G_K)Q_K  - \sum\limits_{a}\sum\limits_{L}\frac{\alpha_{0L}^2}{2\pi^2}T_{aK}C(R_{aL})\times \nonumber\\
&&\times\Big (3C_2(G_L)\ln a_{\varphi\mbox{,}L}-\sum\limits_{b}T_{bL}\ln a_{L}-Q_L\Big(1+\frac{A}{2}\Big)\Big)\nonumber\\
&& +\sum\limits_{a}\sum\limits_{MN}\frac{\alpha_{0M}\alpha_{0N}}{2\pi^2}T_{aK}C(R_{aM})C(R_{aN})\nonumber\\
&& - \sum\limits_{abc}\sum\limits_{L}\frac{\alpha_{0L}}{8\pi^3}T_{aK}C(R_{aL})\lambda_{0i_a m_b n_c}^*\lambda_0^{i_a m_b n_c}\Big(1+A \nonumber\\
&& -B \Big) + \sum\limits_{abc}\sum\limits_{L}\frac{\alpha_{0L}}{4\pi^3}T_{aK}\lambda_{0i_a m_b n_c}^*C(R_{cL})\lambda_0^{i_a m_b n_c}\nonumber\\
&& \times\Big(1+B-A \Big) + \sum\limits_{abc}\frac{\alpha_{0K}}{8\pi^3}T_{aK}C_2(G_K)\lambda_{0i_a m_b n_c}^* \nonumber\\
&& \times \lambda_0^{i_a m_b n_c} - \frac{1}{16\pi^4}\sum\limits_{abcdef}T_{aK}\lambda_{0i_a m_b n_c}^*\lambda_0^{i_a m_b p_d}\nonumber\\
&& \times \lambda_{0p_d k_e l_f}^*\lambda_0^{n_c k_e l_f}\Big\} +O\Big(\alpha_0^3\mbox{,}\alpha_0^2\lambda_0^2\mbox{,}\alpha_0\lambda_0^4\mbox{,}\lambda_0^6\Big).
\label{beta}
\end{eqnarray}

Note that, since the anomalous dimensions (\ref{gamma}) and the \mbox{$\beta$-functions} (\ref{beta}) are defined in terms of the bare couplings, they do not include finite constants specifying a subtraction scheme. As a test of the general expression (\ref{beta}), we calculated with its help the three-loop \mbox{$\beta$-functions} for MSSM regularized by higher covariant derivatives and verified that they coincide with the corresponding expressions found earlier in \cite{Stepanyantz2022}.

The result for three-loop \mbox{$\beta$-functions} standardly defined in terms of the renormalized couplings can be obtained using the renormalization group equations (\ref{beta_0}) and (\ref{beta_r}), and the expression (\ref{beta}). In the particular case where the Yukawa couplings are equal to zero, integrating the equations (\ref{beta_0}) in the two-loop approximation provides the following relations between the bare and renormalized gauge couplings,

\begin{eqnarray}
&&\frac{1}{\alpha_{0K}} - \frac{1}{\alpha_{K}} = \frac{1}{2\pi}\Big\{-Q_K\ln\frac{\Lambda}{\mu}+b_{1,K} + \frac{\alpha_K}{2\pi}C_2(G_K)\times\nonumber\\
&&\times\Big(-Q_K\ln\frac{\Lambda}{\mu} + b_{2,K}\Big) - \sum\limits_{a,L}\frac{\alpha_L}{\pi}T_{aK}C(R_{aL})\times\nonumber\\
&&\times\Big(\ln\frac{\Lambda}{\mu}+b_{2,KL}\Big)\Big\} + O(\alpha^2),
\label{alpha_relation}
\end{eqnarray}
\noindent
where $b_{1,K}$, $b_{2,K}$ and $b_{2,KL}$ are constants of integration that determine a renormalization prescription. The three-loop contributions to the $\beta$-functions defined in terms of the renormalized couplings depend on these integration constants. As a result of the calculation based on the equations (\ref{beta_r}), it was found that in the three-loop approximation

\begin{eqnarray}
&&\frac{\widetilde\beta_{K}(\alpha)}{\alpha^2_K} = -\frac{1}{2\pi}\Big\{-Q_K - \frac{\alpha_K}{2\pi}C_2(G_K)Q_K \nonumber\\
&& - \sum\limits_{a,L}\frac{\alpha_L}{\pi}T_{aK}C(R_{aL})  - \sum\limits_{a,L}\frac{\alpha_K\alpha_L}{2\pi^2}T_{aK}C_2(G_K)C(R_{aL}) \nonumber\\
&& - \frac{\alpha^2_K}{4\pi^2}C_2(G_K)Q_K\Big(C_2(G_K) + b_{2,K} - b_{1,K}\Big) \nonumber\\
&& + \sum\limits_{a,M,N}\frac{\alpha_M\alpha_N}{2\pi^2}T_{aK} C(R_{aM})C(R_{aN}) \nonumber\\
&& - \sum\limits_{a,L}\frac{\alpha^2_L}{2\pi^2}T_{aK}C(R_{aL}) \Big[3C_2(G_L)\ln a_{\varphi,L} - \sum\limits_{b}T_{bL}\ln a_{L}\nonumber\\
&& - b_{1,L} - Q_L\Big(b_{2,KL} + 1 + \frac{A}{2}\Big)\Big]\Big\} + O(\alpha^3).
\label{Renorm_beta}
\end{eqnarray}

From the expression (\ref{Renorm_beta}) it is evident that the terms defining the three-loop contributions to $\widetilde\beta_{K}(\alpha)$ are scheme dependent, with the exception of the terms proportional to $\alpha_M\alpha_N$ for $M\ne N$.

The result for the $\overline{\mbox{DR}}$ scheme can be obtained from the equations (\ref{Renorm_beta}) if the finite constants determining the renormalization prescription are set equal to

\begin{eqnarray}
&& b_{1,K} = 3C_2(G_K)\ln a_{\varphi,K} - \sum\limits_{a}T_{aK}\ln a_K; \nonumber\\
&& b_{2,K} = -\frac{1}{4}Q_K + b_{1,K}; \nonumber\\
&& b_{2,KL} = -\frac{1}{4} - \frac{A}{2}.
\label{DRconditions}
\end{eqnarray}

In this case the expressions for the three-loop $\beta$-functions in the $\overline{\mbox{DR}}$ scheme for a theory with multiple couplings, provided that the Yukawa couplings are absent, take the form

\begin{eqnarray}
&&{\frac{\widetilde\beta_{K}(\alpha)}{\alpha^2_K}}\bigg|_{\overline{\text{DR}}} = -\frac{1}{2\pi}\Big\{-Q_K - \frac{\alpha_K}{2\pi}C_2(G_K)Q_K \nonumber\\
&& - \sum\limits_{a,L}\frac{\alpha_L}{\pi}T_{aK}C(R_{aL}) - \sum\limits_{a,L}\frac{\alpha_K\alpha_L}{2\pi^2}T_{aK}C_2(G_K)C(R_{aL}) \nonumber\\
&& - \frac{\alpha^2_K}{4\pi^2}C_2(G_K)Q_K\Big(-\frac{1}{4}Q_K + C_2(G_K)\Big)\nonumber\\
&& + \sum\limits_{a,L}\frac{3\alpha^2_L}{8\pi^2}T_{aK}C(R_{aL})Q_L \nonumber\\
&& + \sum\limits_{a,M,N}\frac{\alpha_M\alpha_N}{2\pi^2}T_{aK}C(R_{aM})C(R_{aN})\Big\} + O(\alpha^3).
\end{eqnarray}

To verify the correctness of this expression, it was used for calculating contributions to the three-loop MSSM $\beta$-functions that are independent of the Yukawa constants. In the $\overline{\mbox{DR}}$ scheme, the considered $\beta$-functions were first calculated in \cite{Jack2004}. The comparison carried out demonstrated that the contributions under consideration coincided.

\medskip

The author expresses gratitude to K.V. Stepanyants for useful discussions and comments.

\makeatletter\renewcommand\@biblabel[1]{#1.}\makeatother

\end{document}